\documentclass[amssymb,amsmath,prd,twocolumn,superscriptaddress,nofootinbib]{revtex4-1}

\usepackage{graphicx}
\usepackage{bm}
\usepackage{amssymb,amsmath}
\usepackage{mathrsfs}
\usepackage{latexsym}
\usepackage{color}
\usepackage[normalem]{ulem}
\usepackage{dcolumn}
\usepackage[colorlinks=true,citecolor=blue,urlcolor=blue]{hyperref}
\usepackage[usenames,dvipsnames]{xcolor}
\usepackage{xspace}
\usepackage{graphicx}
\usepackage{multirow}
\usepackage{mathtools}

\allowdisplaybreaks

\newcommand{\CIT}{\affiliation{Department of Physics, California Institute of Technology, Pasadena, California 91125, USA}}
\newcommand{\CITLab}{\affiliation{LIGO Laboratory, California Institute of Technology, Pasadena, California 91125, USA}}
\newcommand{\JPL}{\affiliation{Jet Propulsion Laboratory, California Institute of Technology, Pasadena, California 91109, USA}}

\definecolor{shgreen}{rgb}{0.15625, 0.609375, 0.316406}

\newcommand{\neff}{n_{\mathrm{eff}}}
\newcommand{\BF}{\mathcal B^{\mathrm{HD}}_{\mathrm{CP}}}

\newcommand{\gammaGW}{\gamma_{\mathrm{GW}}}
\newcommand{\AGW}{A_{\mathrm{GW}}}

\begin{document}


\title{Accurate characterization of the stochastic gravitational-wave background with pulsar timing arrays by likelihood reweighting}

\author{Sophie Hourihane} 
\thanks{sohour@caltech.edu}
\CIT \CITLab 
\author{Patrick Michael Meyers}
\thanks{pmeyers@caltech.edu}
 \CIT
\author{Aaron Johnson} 
\thanks{aaronj@caltech.edu}
\CIT
\author{Katerina Chatziioannou} 
\thanks{kchatziioannou@caltech.edu} \CIT \CITLab 
\author{Michele Vallisneri}
\thanks{vallis@caltech.edu}
\JPL

\date{\today}

\begin{abstract}
An isotropic stochastic background of nanohertz gravitational waves creates excess residual power in pulsar-timing-array datasets, with characteristic interpulsar correlations described by the Hellings-Downs function.
These correlations appear as nondiagonal terms in the noise covariance matrix, which must be inverted to obtain the pulsar-timing-array likelihood.
Searches for the stochastic background, which require many likelihood evaluations, are therefore quite computationally expensive.
We propose a more efficient method: we first compute approximate posteriors by ignoring cross correlations, and then reweight them to exact posteriors via importance sampling.
We show that this technique results in accurate posteriors and marginal likelihood ratios, because the approximate and exact posteriors are similar, which makes reweighting especially accurate. The Bayes ratio between the marginal likelihoods of the exact and approximate models, commonly used as a detection statistic, is also estimated reliably by our method, up to ratios of at least $10^6$.
\end{abstract}

\maketitle

\section{Introduction}

The nanohertz stochastic gravitational-wave (GW) background  can be detected through the induced delay on the times of arrival of pulses from millisecond pulsars~\cite{Detweiler1979,hellings_upper_1983,Taylor2021}.
Recent evidence that the datasets collected by the three major pulsar-timing-array (PTA) consortia all include excess timing noise of common amplitude and spectral shape~\cite{NANOGrav:2020bcs, Goncharov:2021oub, Chen:2021rqp, Antoniadis:2022pcn} suggests that we might be getting closer to detection~\cite{NANOGRAV:2018hou, pol_forecasting_2022}.
However, since such common-spectrum noise may arise from a non-GW astrophysical or terrestrial source~\cite{Goncharov:2021oub, ZicHobbs2022} (even if this seems unlikely in current data~\cite{NANOGrav:2020bcs,Goncharov:2022ktc}), a GW detection claim needs to wait for the finding that the excess noise is correlated across pulsars with the characteristic angular pattern known as the Hellings-Downs curve~\cite{hellings_upper_1983}.

In PTA data analysis, timing noise is represented as a Gaussian process with covariance matrix $C_{ai \, bj}$, where $a, b$ range over pulsars and $i, j$ over timing measurements (or equivalently frequency components).
For common-spectrum uncorrelated noise, the matrix factorizes as $C_{ij} \delta_{ab}$; for an isotropic GW background, it is given by $C_{ij} \Gamma_{ab}$, with $\Gamma_{ab} = \Gamma(\theta_{ab})$ the Hellings-Downs correlation coefficient, a function of the angular separation $\theta_{ab}$ between pairs of pulsars.
The PTA data model includes several other stochastic components, but GW detection is usually formulated by comparing a common process (CP) model that includes common-spectrum uncorrelated noise and an ``HD'' model that includes common-spectrum Hellings-Downs--correlated noise.\footnote{In the NANOGrav 12.5yr stochastic background analysis that initially reported the evidence for a common process~\cite{NANOGrav:2020bcs}, the CP and HD models are labeled model 2A and 3A, respectively.}
By contrast, information about the GW amplitude and spectral shape is carried primarily by the autocorrelation terms (the $C_{ai \, bj}$ elements with $a = b$).

Parameter estimation and model selection for the CP and HD models are both typically handled through stochastic sampling, which requires repeated evaluations of the data likelihood.
Since the CP excess-noise covariance matrix factorizes across pulsars but the HD matrix does not~\cite{Taylor:2022yeb}, the likelihood is significantly slower to compute for the latter model (e.g., a factor of $\sim 25$ for the NANOGrav 12.5yr dataset, which will only grow larger as more pulsars are observed). 
The number of likelihood evaluations is magnified by the thinning of sample chains (typically by $N_t \sim \mathcal{O}(10^3)$) and by the use of parallel tempering schemes (typically by $N_c \sim \mathcal{O}(10)$ temperatures) which require many likelihood evaluations per CP posterior sample.
The overall cost can be prohibitive for the HD model, particularly when multiple background analyses (e.g., ``sky scrambles''~\cite{Cornish:2015ikx, taylor_all_2017} and ``phase shifts''~\cite{taylor_all_2017}) are required to estimate the significance of a result.

 Methods to optimize PTA search strategies in both data acquisition and modeling have been studied extensively. On the data acquisition side, studies found the most impactful observing cadences and radio frequency bands for detecting a GW background (GWB)~\cite{Lee:2012bf,Lam:2017ysu,Lam:2018uta}. On the modeling side, improvements in computational efficiency have been made by using Fourier basis methods~\cite{LentatiAlexander2013,vanHaasterenVallisneri2014} to characterize red-noise processes, as opposed to dense covariance matrix approaches~\cite{vanHaasteren:2008yh}. More recently, the factorized likelihood approach reduces the wall clock time needed to evaluate a CP-only model by a factor proportional to the number of pulsars~\cite{Taylor:2022yeb}, and Hamiltonian Monte Carlo methods have been implemented to improve sampling efficiency~\cite{Freedman:2022tnf}.

In this study we propose an approach that further mitigates computational cost of producing posterior samples for the HD model in terms of both CPU and wall clock time. Rather than exploring the HD model stochastically, we reuse parameter-estimation results for the inexpensive CP model and ``reweight'' them to obtain posteriors and marginal likelihoods under the HD model. Specifically, a thinned set of CP-model samples yields a set of weighted HD-model samples, with weights equal to the ratios of the HD and CP likelihoods. The computational gains are realized by performing only one HD likelihood evaluation per HD posterior sample, and parallelizing the calculation of weights.

The general reweighting formalism can be applied to any combination of models, though convergence and low sampling error depend on stochastic chains for the original posterior having a sufficient number of samples in the support of the target posterior. This is the case for the HD and CP posteriors, since both are dominated by single-pulsar autocorrelation terms. Additionally, the two models share the same parameters and corresponding priors.
In this paper we apply the reweighting formalism to simulated PTA data, and compare posteriors and marginal likelihoods obtained by reweighting and by brute-force sampling. We find that (i) the posteriors recovered through reweighting are statistically unbiased; and that (ii) the HD vs.\ CP Bayes factors (the ratios of marginal likelihoods) agree with the ``hypermodel'' method typically used in PTA analyses~\cite{hee_bayesian_2016} to within 10$\%$ uncertainty for Bayes factors $\in [10^{-3}, 10^7]$.

The rest of the paper is organized as follows. In Sec.~\ref{sec:formalism} we introduce the general reweighting formalism following~\cite{Payne:2019wmy}. In Sec.~\ref{sec:models} we describe the HD and CP models in more detail. In Sec.~\ref{sec:results} we present results from simulated data that validate the reweighting approach. In Sec.~\ref{sec:discussion} we conclude by discussing the application of our method and its computational gains.

\section{Posterior reweighting}
\label{sec:formalism}

Samples distributed according to one posterior distribution can, under some circumstances, be reweighted to estimate a second posterior distribution; this is a form of importance sampling.
In this section, we present the general methodology behind this posterior reweighting following Ref.\ \cite{Payne:2019wmy}, and describe how it can be used to also estimate the marginal likelihood of a model and the Bayes factor between models.

The posterior distribution, $p(\theta | d, T)$ for a target model $T$ with parameters $\theta$ given data $d$ can be written explicitly in terms of the Bayes theorem,
\begin{align}
    p(\theta | d, T) = \frac{\mathcal{L}(d | \theta, T)\pi(\theta | T)}{\mathcal Z_{T}}\,,
\end{align}
where $\mathcal{L}(d | \theta, T)$ is the likelihood, $\pi(\theta | T)$ is the prior, and $\mathcal Z_{T}$ is the marginal likelihood (also known as evidence, though we do not use this term here). We rewrite this target posterior distribution in terms of the likelihood and prior for another ``approximate'' model $A$,
\begin{align}
    p(\theta | d, T) &= \frac{\mathcal{L}(d | \theta, A)\frac{\mathcal{L}(d | \theta, T)}{\mathcal{L}(d | \theta, A)}\pi(\theta | A)\frac{\pi (\theta | T)}{\pi(\theta | A)}}{\mathcal Z_{T}}\\
    &=w_{\mathcal L}(d|\theta)w_{\mathcal \pi}(\theta)\frac{\mathcal{L}(d | \theta, A)\pi(\theta | A)}{\mathcal Z_{T}}\,.
\end{align}
In the last line we have introduced weights given by the ratio of the likelihoods and priors of the two models 
\begin{align}
    w_{\mathcal L}(d|\theta) &= \frac{\mathcal{L}(d | \theta, T)}{\mathcal{L}(d | \theta A)}\,,\label{eq:likelihoodWeights}\\
    w_{\mathcal \pi}(\theta) &= \frac{\pi (\theta | T)}{\pi(\theta | A)}; \label{eq:priorWeights}
\end{align}
we can also combine the weights to get
\begin{align}
    w(d|\theta) =  w_{\mathcal L}(d|\theta) w_{\mathcal \pi}(\theta) \,.
\end{align}

Given $N_s$ posterior samples $\theta_s \sim p(\theta|d,A)$ for model $A$, we can resample them with weights $w(d|\theta_s)$ to obtain a posterior sampling of model $T$; the marginal likelihood $\mathcal Z_T$ can also be estimated as
\begin{align}
    \mathcal Z_T &= \int d\theta \, \mathcal{L}(d | \theta, T)\pi (\theta | T)\\
    &=\mathcal Z_{A} \int d\theta\, w_{\mathcal L}(d|\theta)w_{\mathcal \pi}(\theta) p(\theta | d, A)\,.\label{eq:evidence_reweighting}
\end{align}
The integral in Eq.~\eqref{eq:evidence_reweighting} can be approximated with Monte Carlo integration:
\begin{align}
    \mathcal Z_{T} \approx \frac{\mathcal Z_A}{N_s}\sum_{s=1}^{N_s} w_{\mathcal L}(d|\theta_s)w_{\mathcal \pi}(\theta_s) = {\mathcal Z_A} \bar{w}\,,
    \label{eq:evidence_reweighting_monte_carlo}
\end{align}
where $\bar{w}$ is the mean of the weights, $w(d|\theta)$.
If we are interested in model selection between the approximate and target models, the Bayes factor between them is then simply
\begin{align}
\mathcal B^{T}_{A} = \bar{w}\,.\label{eq:BayesFactor}
\end{align}

Though the reweighting procedure is mathematically exact, it is subject to sampling errors, especially if the approximate and target posteriors are too disjoint.
We quantify sampling error with the ``effective number of samples'' $\neff$---the approximate number of samples drawn independently from the target posterior that would approximate $\mathcal{Z_T}$ as accurately as the reweighting estimate \eqref{eq:evidence_reweighting_monte_carlo}. Reference \cite{Elvira_2022} estimates $\neff$ as
\begin{align}
    \neff &\approx \frac{\left[\sum_s w_{\mathcal L}(d|\theta_s)w_{\mathcal \pi}(\theta_s)\right]^2}{\sum_s \left[w_{\mathcal L}(d|\theta_s)w_{\mathcal \pi}(\theta_s)\right]^2}\, 
    = \frac{N_s}{1 + \left(\frac{\sigma_w}{\bar{w}}\right)^2}\,,\label{eq:neff}
\end{align}
where $\sigma_w$ is the standard deviation of the weights. 
We also define the efficiency
\begin{align}
    \mathcal{E} \equiv \frac{\neff}{N_s}\,.\label{eq:Efficiency}
\end{align}
It follows from Eq.~\eqref{eq:BayesFactor} that the error $\sigma_{\mathcal{B}}$ on the mean $\mathcal B^{T}_{A}$  is
\begin{align}
    \sigma_{\mathcal{B}} = \frac{\sigma_w}{\sqrt{\neff}} = \frac{\sigma_w}{\sqrt{\mathcal{E}  \, N_s} } \,.\label{eq:BayesFactorError}
\end{align}

If we represent the target posterior by a set of equal-weight samples by performing a weighted redraw from the approximate distribution, then Eq.~\eqref{eq:neff} makes intuitive sense. It implies that a few samples with high weights (relative to $\bar{w}$), will result in the same sample being drawn many times and lead to comparatively lower $\neff$. Equivalently, such high individual weights (relative to $\bar{w}$) increase $\sigma_w$ and thus decrease $\neff$. In the limit of vanishing variance, $\neff \rightarrow N_s$, while as variance grows $\neff \rightarrow 0$.

We can also use the weights to estimate the statistical distance between the approximate and target posteriors in the form of the Kullback-Leibler (KL) divergence. That is, 
\begin{align}
    \mathcal{D}_{KL}(A|| \,T) &\equiv 
    \int d\theta \, p_T(\theta) \ln{\frac{ p_A(\theta)}{ p_T(\theta)}} \,, \\
    &\approx \sum\limits_{\theta \sim p_A} \frac{1}{N_s} \ln{\frac{ p_A(\theta)}{ p_T(\theta)}} \, , \\
    &= \ln(\bar{w}) - \overline{\ln(w)}\,, \label{eq:KLDivergence}
\end{align}
where we have written the posteriors $p_K(\theta) \equiv p(\theta | d, K)$ for $K\in\{A, T\}$ and $\overline{\ln(w)}$ is the average of the log of the weights. This equation can be used in combination with Eq.~\eqref{eq:neff} to provide guidance when reweighting leads to low efficiency.

The main reason for low efficiency is that a region of low posterior for the approximate model overlaps with a region of high posterior for the target model. Samples in this region get very high weights and can lead to poor reconstruction of the target posterior. Understanding when and where this can occur is an area of active research, and has led to other forms of importance sampling~\cite{vehtari_reweighting}.

\section{The pulsar-timing-array models for stochastic gravitational waves}
\label{sec:models}

In this section we discuss the statistical framework used to detect a stochastic GW background with an array of regularly timed millisecond pulsars. We first introduce a Gaussian likelihood that includes the full interpulsar correlations induced by GWs (the Hellings-Downs model).
We then introduce a secondary model that ignores interpulsar correlations and includes GWs as a common (but uncorrelated) power law spectrum in each pulsar’s residuals (a common process model).
We claim evidence for GWs when a dataset significantly favors HD over CP.

These two models contain the same parameters and priors; furthermore, the posterior distributions of model parameters are not affected strongly by the inclusion of interpulsar correlations. This makes the CP likelihood a good approximate distribution for the HD likelihood. As we discuss below, the CP model is significantly faster to evaluate, and so we will use CP as our approximate likelihood, $\mathcal L_{A}$, while the HD model will be the target $\mathcal L_{T}$. An added bonus of our choice of these models is that, in implementing the reweighting scheme discussed above to speed up computation of the HD posterior, we naturally also calculate Bayes factors that can be used as a GW detection statistic.

\subsection{Pulsar-timing-array likelihood}
\label{sec:likelihood}

A detailed presentation of the PTA likelihood derivation can be found in Refs.~\cite{vanHaasterenLevin2013,LentatiAlexander2013,vanHaasterenVallisneri2014,ArzoumanianBrazier2016,taylor_all_2017,Taylor2021}; in this section we describe only the relevant details.  A reader familiar with PTA analyses can skip to Sec.~\ref{sec:computationalCost}.

Pulse arrival times [time(s) of arrival (TOA)] are affected by both deterministic and stochastic processes. The deterministic contribution (described more fully in~\cite{taylor_all_2017}) contains terms relating to the motion of the pulsar, such as sky location, rotation period, etc., as well as individually resolvable GW sources such as continuous waves. An initial solution for the timing model is subtracted from the measured TOA, leaving behind the fit residuals $\delta \bm{t}$. The uncertainties on this timing model are described by a Taylor expansion in timing parameters $\bm{\epsilon}$ with partial derivatives (comprising the design matrix) $\bm{M}$ evaluated at the initial timing solution. 
 
The stochastic contribution to the TOA is due to a combination of the intrinsic low-frequency spin noise (or ``red noise'') of individual pulsars (IRN) and a common, stochastic process induced by a GW background. We model both as stationary zero-mean Gaussian random processes with Fourier vector bases. It follows that the Fourier coefficients $\bm{a}$ (the basis weights) are described entirely by their covariance $\bm{\phi}|_{\bm{\eta}} = \langle \bm{a}_{ai} \bm{a}_{bj} \rangle$. Here indices $a,b$ index pulsars, $i,j$ index frequencies, brackets indicate the ensemble average, and the $\bm{\eta}$ are the ``hyperparameters" associated with the distribution of $\bm{a}$.
 
With both the deterministic and stochastic contributions to the noise modeled, the timing residuals $\bm{r}$ are
\begin{align}
    \bm{r} = \delta \bm{t} - \bm{M}\bm{\epsilon} - \bm{F} \bm{a} =  \delta \bm{t} - \bm{T} \bm{b}\,,
\end{align}
where the matrix $\bm{F}$ collects the Fourier basis vectors evaluated at the TOA and where we have introduced
\begin{align}
     \bm{T} \equiv [\bm{M} \, \bm{F}]\,,\quad \bm{b} &\equiv \begin{bmatrix}
           \bm{\epsilon} \\
           \bm{a}
         \end{bmatrix}\,,
\end{align}
for ease of notation. 
The residuals $\bm{r}$ should now be white and Gaussian, with a covariance matrix $\bm{N}$ that describes the uncertainty associated with each TOA observation. The likelihood is then 
\begin{align}
    \mathcal{L}(\delta \bm{t} | \bm{b}) =\frac{ \exp \left(-\frac{1}{2} \bm{r}^{T} \bm{N}^{-1}\bm{r}\right)}{\sqrt{\det{2 \pi \bm{N}}}}\,.\label{eq:unmargenalizedLikelihood}
\end{align}
We complement the likelihood with the Gaussian-process prior for the Fourier components,
\begin{align}
    \pi(\bm{a} |\bm{\eta}) = \frac{\exp(-\frac{1}{2} \bm{a}^T \bm{\phi}|_{\bm{\eta}}^{-1} \bm{a})}{\sqrt{\det{2\pi \bm{\phi}|_{\bm{\eta}}}}}\,. \label{eq:GaussianPrior}
\end{align}
The Gaussian form of the likelihood and prior means that we can marginalize analytically over the $\bm{a}$, leaving only the hyperparameters $\bm{\eta}$. A similar choice is made for the timing model correction prior $\pi(\bm{\epsilon})$ \cite{taylor_all_2017}. The marginalized likelihood is then
\begin{align}
    \mathcal{L}(\delta \bm{t} | \bm{\eta}) &= 
    \int d\bm{b} \; \mathcal{L}(\delta \bm{t} | \bm{b}) \; \pi(\bm{a}|\bm{\eta}) \; \pi(\bm{\epsilon}) \,, \\
    &\propto \frac{\exp\left(-\frac{1}{2} \delta \bm{t}^T \bm{C}^{-1}\delta \bm{t} \right)}{\sqrt{\det\left(2\pi \bm{C} \right)}}\,,\label{eq:likelihood}
\end{align}
where $\bm{C} = \bm{N} + \bm{T}\bm{B}\bm{T^T}$ is the covariance kernel, and 
\begin{align}
    \bm{B} = 
    \begin{bmatrix}
    \bm{\infty} & 0 \\
    0 & \bm{\phi} 
    \end{bmatrix}\,.
\end{align}
Here $\bm{\infty}$ represents a formal limit of covariance for a uniform unbounded prior on $\bm{\epsilon}$.  

\subsection{Pulsar-timing-array stochastic models}
\label{sec:stochastic_models}

We model both the IRN and the GWB as power laws in the frequency domain.\footnote{Power laws are not the only choice for the distribution of Fourier coefficients. Other choices include (but are not limited to) a free spectral model with independent densities for each Fourier frequency, and a broken power law~\cite{NANOGrav:2020bcs, Taylor:2016ftv, Sampson:2015ada}.} The model hyperparameters are then $\bm{\eta} = (A^a, \gamma^a, \AGW, \gammaGW)$ where $\gamma^a$, $\gammaGW$ and  $A^a$, $\AGW$ are the negative spectral indices and amplitudes of the IRN and GW power laws respectively. We split $\bm{\phi}$ into its two contributions; one from the IRN and the other the common GWB
\begin{align}
    \bm{\phi} = \bm{\phi}^{\textrm{IRN}} + \bm{\phi}^{\mathrm{GW}}\,.\label{eq:PhiTot}
\end{align}
By stationarity, both the IRN and the GWB are uncorrelated between frequencies. Therefore $\phi$ will contain no cross-frequency terms, and $\bm{\phi}_{ai, bj} \propto \delta_{ij}$.

By definition, the IRN is an independent process in each pulsar:
\begin{align}
    \bm{\phi}|_{\bm{\eta}\;\;(ai),(bj)}^{\textrm{IRN}} = \kappa_{i}(\bm{\eta}_a)\delta_{ab}\delta_{ij}\,. \label{eq:phiIntrinsicRed}
\end{align}
The IRN power $\kappa_{i}(\bm{\eta}_a)$ in frequency bin $i$ for pulsar $a$ is modeled as the power law
\begin{align}
    \kappa_{i}(\bm{\eta}_a) = \kappa_{i}(A_a, \gamma_a) = \frac{A_a^2}{12\pi^2} \frac{1}{T} \left( \frac{\nu_i}{\mathrm{yr^{-1}}}\right)^{-\gamma_a}\mathrm{yr^2}\,,\label{eq:kappa_powerlaw}
\end{align}
where $T$ is the total observation time and $\nu_i$ is the frequency associated with bin $i$. 

In contrast to the IRN, the GW background is correlated between pulsars:
\begin{align}
    \bm{\phi}|_{\bm{\eta}\;\;(ai),(bj)}^{\textrm{GW}}  = \Gamma_{ab} \kappa_{i}(\bm{\eta}_{\mathrm{GW}})\delta_{ij}\,. \label{eq:phiGW}
\end{align}
Here $\kappa_i$ is again given by Eq.~\eqref{eq:kappa_powerlaw}, except that every pulsar has the same amplitude $\AGW$ and spectral index $\gammaGW$. The function $\Gamma_{ab}$ describes GW correlations between pulsars $a$ and $b$ and is known as the Hellings-Downs curve [\cite{hellings_upper_1983}, Eq. (5)]. 

The $a = b$ components of Eq.~\eqref{eq:phiGW} represent the excess-noise power induced by the GWB in each pulsar. Half of this power is caused by the ``Earth term" in the pulsar GW response, and contributes to interpulsar correlations; the other half is caused by the ``pulsar term" and is statistically uncorrelated among pulsars. 
The first indications of a GWB in PTA data will appear through these diagonal self-correlations \cite{NANOGRAV:2018hou,Romano:2020sxq}, so they could be detected using the CP model as well as the HD model.
However, evidence for CP could also be caused by physical effects such as the solar wind~\cite{Tiburzi:2015kqa} or by model misspecification, such as incorrect priors~\cite{ZicHobbs2022} or poor IRN models~\cite{Goncharov:2022ktc}.
Other mechanisms can induce interpulsar correlations that are inconsistent with the Hellings-Downs curve, such as clock errors with monopolar correlations \cite{Tinto:2018bae, Tiburzi:2015kqa} or Solar System ephemeris errors with dipolar correlations~\cite{Goncharov:2020krd, NANOGrav:2020tig_bayesephem, Caballero:2018lvc, Roebber:2019gha}.
Thus, the detection of HD correlations through the off-diagonal terms of $\phi_{\mathrm{GW}}$ is considered the decisive factor in claiming a GWB detection, and the CP vs. HD Bayes factor is used as a GWB detection statistic~\cite{NANOGrav:2020bcs, Chen:2021rqp, Goncharov:2022ktc, Antoniadis:2022pcn}.

\subsection{Implementation and computational considerations} \label{sec:computationalCost}

\begin{figure}
    \centering
    \includegraphics[width=.5\textwidth]{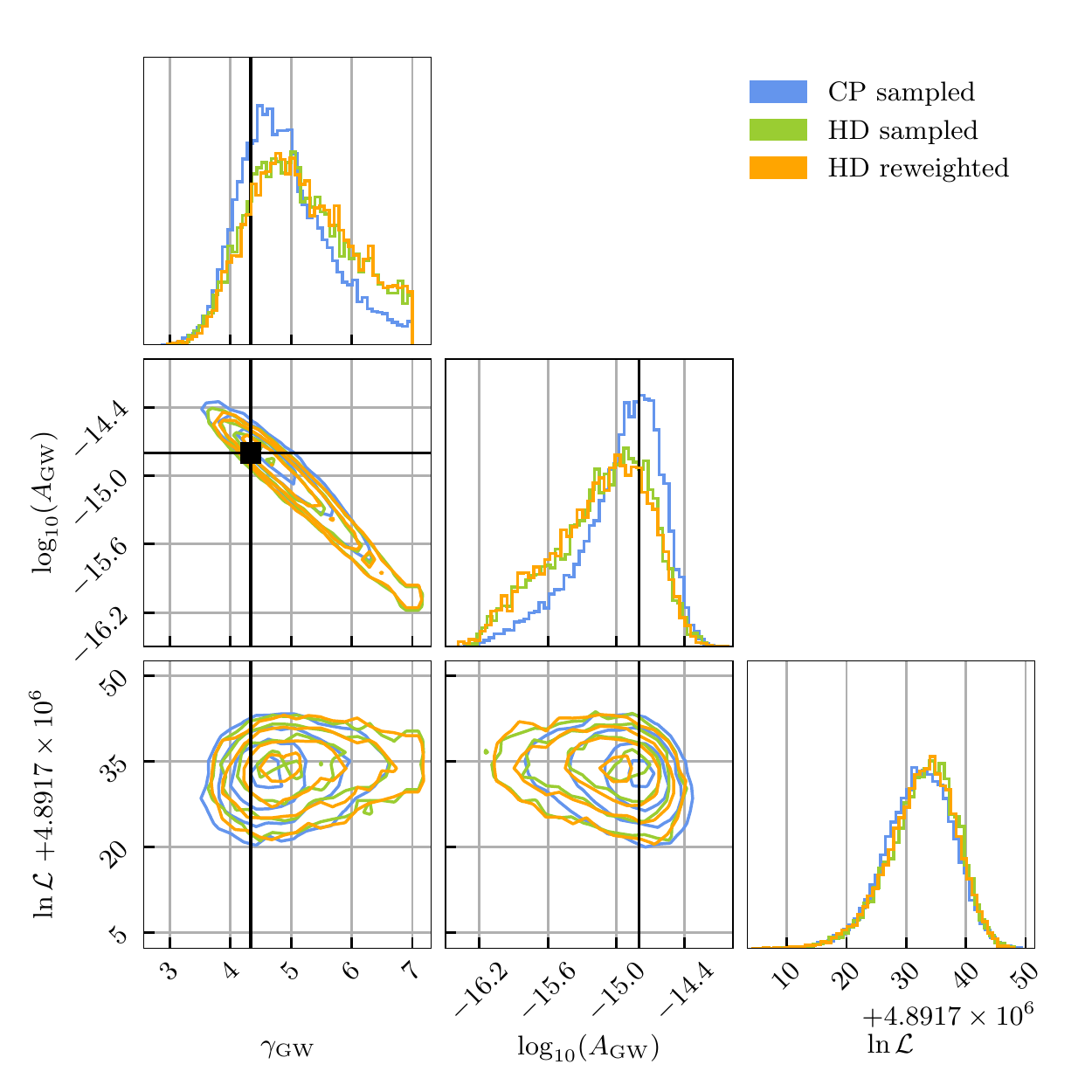}
    \caption{Posteriors for $\gammaGW$ and $\log_{10}\AGW$ and $\ln{\mathcal{L}}$ distribution for simulated PTA data with a $\log_{10}(A_\mathrm{GW}) = -14.8$ GWB.
    We show histograms for direct sampling of CP (blue), HD (green), and for CP-to-HD reweighting (orange). 
    Black lines indicate the injected values.
    For this plot we selected one of our simulations with the most visually different CP and HD GW posteriors. Even so, the direct-sampling and reweighted HD posteriors are almost identical. The reweighted posterior is well sampled, with 51$\%$ efficiency.} 
    \label{fig:m2a_vs_m3a_posteriors}
\end{figure}

The standard PTA likelihood \eqref{eq:likelihood} requires the inverse noise covariance matrix $\bm{C}^{-1}$ and therefore the inverse of $\bm\phi|_{\bm\eta}$. Although the PTA analysis software, such as \textsc{Enterprise}~\cite{enterprise}, is optimized to speed up the likelihood evaluation, inversion becomes the most expensive computation when $\bm{\phi|_\eta}$ is not pulsar diagonal.
For instance, for the NANOGrav ``12.5yr'' dataset each HD-model likelihood is $\sim 25$ slower than the corresponding CP likelihood. This factor applies to 45 pulsars over a 12.9 year dataset and will increase with the number of pulsars.

The current workhorse method to compute $\BF$ is a hypermodel Markov chain Monte Carlo sampler~\cite{hee_bayesian_2016, justin_ellis_2017_1037579, Taylor:2020zpk}. In such an analysis, a discrete metaparameter tracks the current model (HD or CP) while the sampler jumps between them. The final Bayes factor is the number of samples in the HD model divided by the number of samples in the CP model. The two posteriors are also selected by the value of the metaparameter.
As the evidence for a GWB becomes stronger, the HD model will be sampled more often than the CP model, slowing down calculations further.\footnote{A constant added to the CP log-likelihood can mitigate this particular issue and result in a comparable number of samples in each model. That constant should be close to the Bayes factor, which is unknown \emph{a priori} in real data. In our study we estimated this constant by using the likelihood ratio between the CP and HD models evaluated at the injected parameters. This ensured that both models contained enough samples particularly in high Bayes factor regimes.}

Despite the difference in likelihood functions and computation time, the posteriors for CP and HD are generally quite similar. Figure~\ref{fig:m2a_vs_m3a_posteriors} displays the marginalized one- and two-dimensional $\gammaGW$ and $\AGW$ posteriors and the $\ln \mathcal{L}$ distributions for CP (blue) and HD (green), as recovered by hypermodel sampling.
The similarity between the posteriors and the $\sim$25x likelihood speedup suggest that this problem is well suited for the reweighting method introduced in Sec.~\ref{sec:formalism}.\footnote{Here we use identical priors between the target and the original distribution, meaning the prior weights $w_{\pi}(\theta) = 1$.}
The HD posterior created by reweighting the CP posterior is plotted in orange and is almost identical to the direct-sampling HD posterior. The efficiency of the reweighting method as posterior differences is discussed in the next section.

\section{Demonstration of the method}
\label{sec:results}

To show that we can safely reweight CP to HD, we simulate PTA datasets containing GWBs with different amplitudes, and demonstrate that reweighting yields unbiased Bayes factors and posteriors.

\subsection{Bayes factors}\label{sec:results_bayesfactors}
\begin{figure}
    \centering
    \includegraphics[]{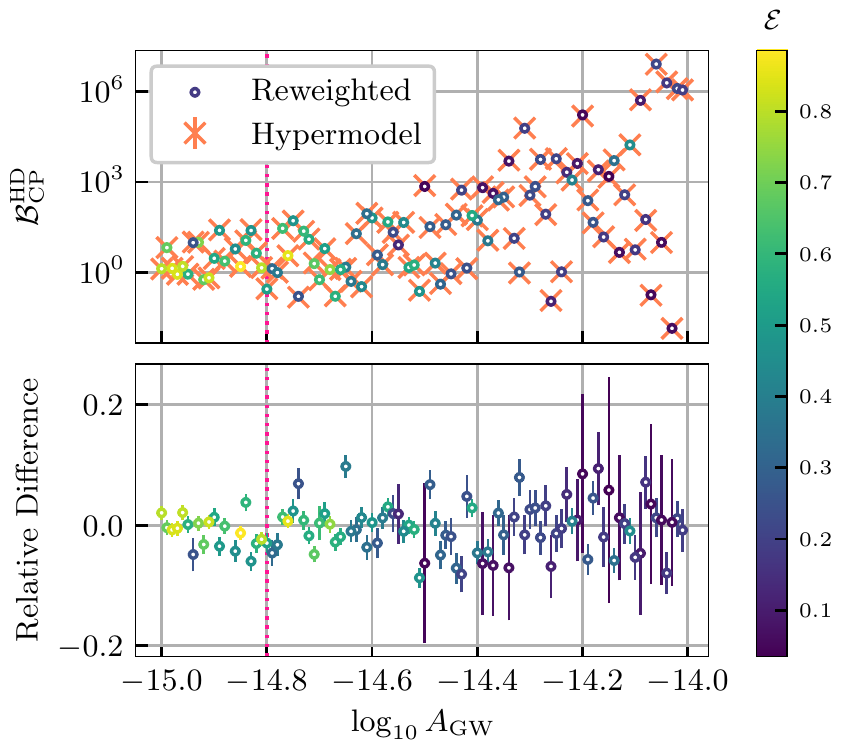}
    \caption{
    Top: $\BF$ vs simulated GWB amplitude. Bayes factors recovered via reweighting, Eq.~\eqref{eq:BayesFactor}, are colored by their efficiency $\mathcal{E}$, Eq.~\eqref{eq:Efficiency}. Bayes factors recovered via the hypermodel are plotted as coral $X$s. The hypermodel error is calculated with a bootstrap method described in~\cite{Heck_2018} whereas the reweighting error is estimated with Eq.~\eqref{eq:BayesFactorError}, although both errors are too small to see.
    Bottom: relative difference in the hypermodel and  recovered Bayes factors, again colored by efficiency. The error bars are propagated from the hypermodel and reweighting errors above. 
    As the GWB amplitude increases, the efficiency decreases due to the distribution of the weights broadening as in Eq.~\eqref{eq:neff}. The relative difference between these Bayes factors is usually small, typically $-0.5 \pm 4 \%$, but can be as large as 10\%. A 10\% difference in $\BF$ is not large enough to change a detection conclusion to a nondetection conclusion or vice versa and therefore we can consider the difference small. For instance, a Bayes factor of 100 would lead to the same qualitative conclusion as a Bayes factor of 110. The pink vertical line in both plots is $\log_{10} A_{\rm GW} = -14.8$, the posterior plotted in Fig.~\ref{fig:m2a_vs_m3a_posteriors} to demonstrate that this posterior is typical.}
    \label{fig:error_BF}
\end{figure}

To test $\BF$ recovery, we simulate 100 datasets for 45 pulsars over 12.9 years, using maximum-likelihood red-noise hyperparameters from the 12.5yr NANOGrav dataset~\cite{NANOgrav:12p5yrData, NANOGRAV:2018hou}.\footnote{The NANOGrav 12.5yr dataset is actually 12.9 years in length.}
Each simulation includes a power-law GWB with $\log_{10}\AGW$ varying uniformly between $-15$ and $-14$.
We set $\gammaGW$ to $13/3$, the theoretical value for a GW background from supermassive black-hole binaries~\cite{Phinney:2001di}.
For each simulated dataset, we obtain a thinned set of CP posterior samples using \textsc{PTMCMCSampler}~\cite{justin_ellis_2017_1037579}. We reweight the CP posterior sample to the HD model and calculate $\BF$ following Eq.~\eqref{eq:BayesFactor}. To verify the accuracy of these reweighted $\BF$s, we obtain an independent estimate from hypermodel runs on the same simulations. 
We compare the reweighted and hypermodel Bayes factors in Fig.~\ref{fig:error_BF}, finding them in excellent agreement. The top panel shows the $\BF$ estimates plotted against $\log_{10} \AGW$; the bottom panel shows the relative difference of the $\BF$ estimates (reweighted minus hypermodel, divided by their average).
Marker colors encode reweighting efficiency.
The mean relative difference is $-0.5\pm4\%$, so we observe no systematic effect.
The maximum relative difference is 10\%, small enough that it could not affect a GWB detection claim.
Error bars are computed by combining (in quadrature) reweighted Bayes factor errors from Eq.~\eqref{eq:BayesFactorError} and hypermodel Bayes factor errors from the bootstrap method of \cite{Heck_2018}. 
Bayes factor differences are not strongly correlated with the injected GW amplitude or the Bayes factor, although the difference uncertainties are inversely correlated with efficiency [see Eq.~\eqref{eq:BayesFactorError}]. 

Figure~\ref{fig:error_BF} shows also that as we increase the simulated amplitude, the sampling efficiency tends to decrease. This is expected; as the amplitude of the GWB increases, the off-diagonal terms in Eq.~\eqref{eq:phiGW} become more significant. The likelihood can then change between the two models significantly, which affects $\bar{w}$, and can even be maximized in different parts of parameter space. Such conditions can lead to a large spread in the weights as some points get heavily upweighted and others get downweighted. From Eq.~\eqref{eq:neff}, a large spread in the weights means that $\neff$ will decrease, and more samples from the CP distribution will be needed in order to faithfully represent the HD posterior and calculate $\BF$ accurately: see Eq.~\eqref{eq:BayesFactorError}. In our simulated datasets, however, the recovered Bayes factor remains within 10\% of that calculated with the hypermodel even in regions where $\BF > 10^6$.

\begin{figure}
    \centering
    \includegraphics[]{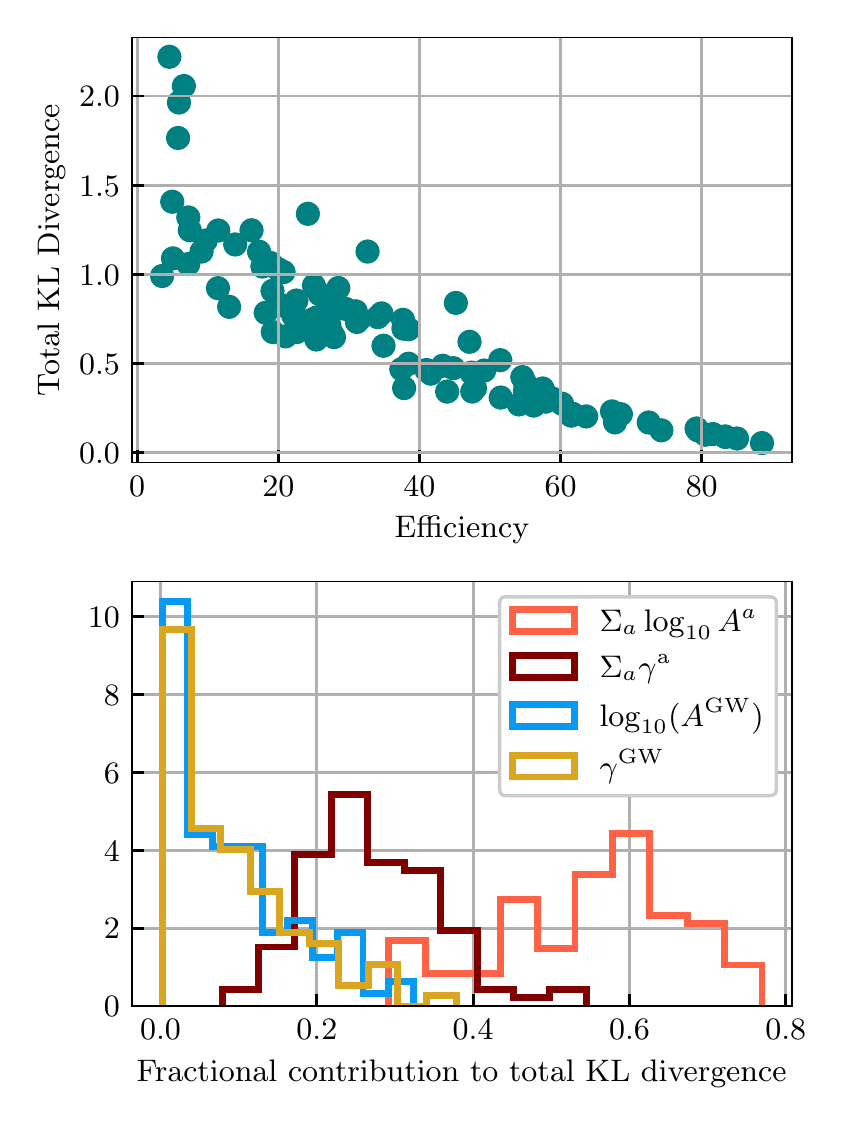}
    \caption{
    Top: total KL divergence, Eq.~\eqref{eq:KLDivergence}, vs efficiency, Eq.~\eqref{eq:Efficiency}, between the CP and the reweighted HD posterior. As the KL divergence increases, the posteriors become more distinct, and the sampling efficiency decreases. 
    Bottom: fractional contributions to total KL divergence from sets of parameters including all the IRN amplitudes and spectral indices (pink and red, respectively) and the GWB amplitude and spectral index (blue and gold respectively). The set of all IRN parameters contribute more to the total divergence than the GWB parameters individually.}
    \label{fig:KLDivergence}
\end{figure}

In order to study the relation between the model posterior similarity and the efficiency of the reweighting procedure, we compute the Kullback–Leibler divergence~\cite{Kullback_1951}, which quantifies the difference between two distributions. We plot the relationship between the KL divergence and the efficiency in Fig.~\ref{fig:KLDivergence}. The upper plot shows total KL divergence [Eq.~\eqref{eq:KLDivergence}] vs.\ efficiency [Eq.~\eqref{eq:Efficiency}] for the CP and HD posteriors. As the KL divergence increases, the posteriors become more distinct and the sampling efficiency decreases.
The bottom plot shows the fractional contributions of different model parameters to the total KL divergence.\footnote{The total KL divergences are not directly comparable to the KL divergences of the various marginalized posteriors; the total KL divergence is equal to the sum of the marginalized KL divergences only when parameters are uncorrelated. Although this is not the case in our analysis, the normalized marginal KL divergences still inform us of parameters that most greatly influence the total KL divergence.}
We split the 92 parameters into four sets: the IRN amplitudes and spectral indices (pink and red respectively) and the GW background amplitude and spectral index (blue and gold respectively). We compute the partial KL divergence of the CP and HD marginalized posteriors for each parameter and sum those of the IRN parameters. The fractional contribution is then obtained by dividing those partial KL divergences by the total.
The set of all red-noise parameters contributes more to the total divergence than do the GWB parameters individually. The set of all IRN amplitude posteriors is the major contributor to the divergence ($55 \pm 11\% $), followed by the set of all IRN spectral indices ($27 \pm 8 \%$); the contribution from $\AGW$ and $\gammaGW$ are roughly equivalent at percent level, $9\pm 8 \%$ each.

\subsection{Posterior recovery}

Figure~\ref{fig:m2a_vs_m3a_posteriors} offered visual confirmation that the GWB parameter posteriors under the HD model are recovered without bias via reweighting. In this section we confirm these initial findings through a more extensive percentile-percentile (P-P) test~\cite{Gibbons2003}. We generate 100 simulations similar to those described in Sec.~\ref{sec:results_bayesfactors}, except that each simulated parameter is drawn from its analysis prior, as required to achieve Bayesian coverage.
The priors for the spectral indices are $\gammaGW, \gamma^{a} \in \mathrm{U}\,[2, 6]$, and the priors for the amplitudes are $\log_{10}\AGW\in \mathrm{U}\,[-15, -12]$ and $\log_{10}A^{a} \in \mathrm{U}[-16, -14]$.
We recover CP posteriors from these simulations with direct sampling, and then reweight and resample those posteriors to the HD model.

\begin{figure}
    \centering
    \includegraphics[]{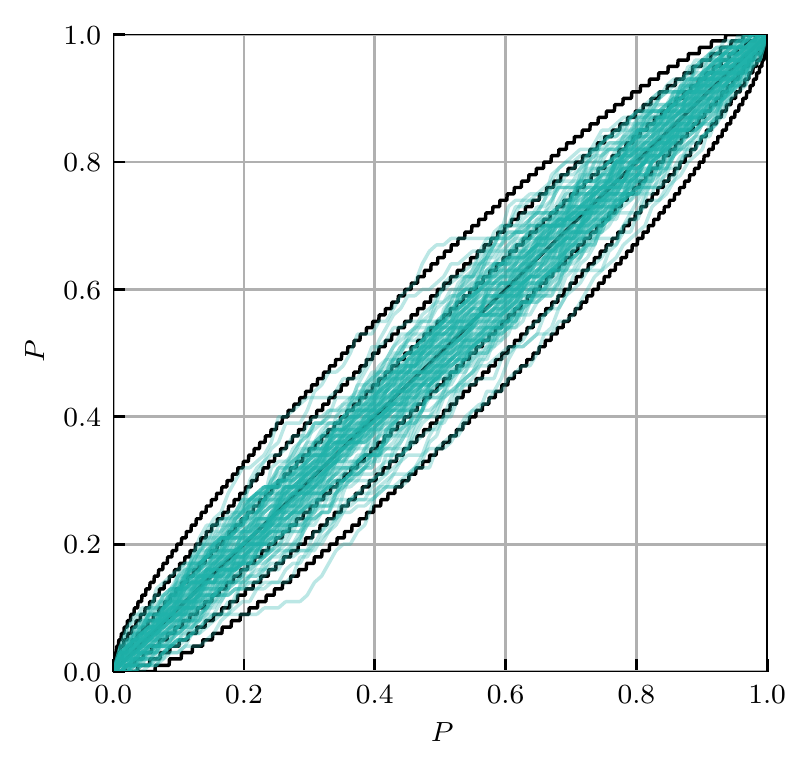}
    \caption{P-P plots for all 92 reweighted, HD-model parameters (teal) with the 1-, 2- and 3-$\sigma$ standard deviations (black). The y-axis is the percentile of each parameter's injected value in its marginalized posterior. The x-axis is the percentile of the sorted y-axis values. The recovered posteriors are consistent with expectations, suggesting that the posterior recovery is unbiased.}
    \label{fig:pp_plot}
\end{figure}

The P-P test is a standard measure of bias in recovered posteriors. Data-sets are first simulated by drawing parameters from their priors and adding Gaussian noise. The posterior of each data-set is then sampled. The percentile of each of the ``true" or injected values is calculated in the marginalized, one dimensional posterior of each parameter. For a set unbiased posteriors, the injected value will be distributed according to each posterior. That is, the percentile of where each injected value lands in a 1-D marginalized posterior will be distributed uniformly between the 0th and 100th percentile, the x-axis of Fig.~\ref{fig:pp_plot}. This test of uniformity in posterior space is represented with the cumulative distribution function (CDF) of the posteriors percentile. Since the CDF of a uniform distribution between 0 and 1 (0th and 100th percentile) is a line of slope 1, the P-P plot is usually represented this way. A P-P plot showing a line consistently below (above) the line x = y is indicative of parameter bias of overestimating (underestimating) the parameter value. An S-curve going above (below) then below (above) the diagonal is indicative of a overestimate (underestimate) of the posterior's standard-deviation.

Figure~\ref{fig:pp_plot} shows the corresponding P-P plots. The 92 different parameters ($\gamma^a, A^a$ for 45 pulsars as well as $\gamma^\mathrm{GW}, A^\mathrm{GW}$ for the GWB) are plotted in teal. The expected 1-, 2- and 3-$\sigma$ confidence intervals are plotted in black. The recovered posteriors agree with expectations; only two lines briefly leave the three-sigma error bars. This suggests that the reweighting method neither over- nor underestimates parameters systematically, as would be the case if some parameters were always above or below the diagonal; nor does it recover incorrect variance, as would be indicated by S-curves around the diagonal.

\subsection{Bayes factor recovery on extended dataset}

To this point, we have demonstrated that likelihood reweighting is a promising tool for recovering accurate Bayes factors and unbiased posteriors in a simulated data-set with 12.9 years of timing data, 45 pulsars, and a range of injected GWB amplitudes. As PTAs continue to collect more data, it becomes natural to ask at what point the reweighting scheme could fail, either by misestimating Bayes factors or by exhibiting low efficiencies. We examine the performance of likelihood reweighting after the addition of additional pulsars and additional observation time to the dataset. We find that even for 80 pulsars and 22 years of data, the efficiency remains above $20\%$ and the errors between the Bayes factor calculated with direct sampling and reweighting are comparable to those in Sec.~\ref{sec:results_bayesfactors}.

To create this extended dataset, we simulate realistic pulsars and add additional observing time to each pulsar. To create new pulsars, we sample sky locations by fitting existing pulsar locations with a kernel density estimate and sample from it. Each new pulsar is assigned white noise parameters and observing epochs (plus Gaussian scatter) from an existing pulsar. To simulate additional years of data, to each pulsar we add TOAs with Gaussian scatter. The red noise parameters for the new pulsars are drawn from the IRN prior. For existing pulsars, the red-noise amplitudes were set from the maximum-likelihood draw as in Sec.~\ref{sec:results_bayesfactors}. The GWB was injected with $\AGW=1.92\times 10^{-15}$, the median posterior amplitude of the NANOGrav 12.5yr analysis~\cite{NANOGrav:2020bcs}. In total, we simulated 22 years of data in 80 pulsars; below we present results based subsets of that data.

\begin{figure}
    \centering
    \includegraphics[]{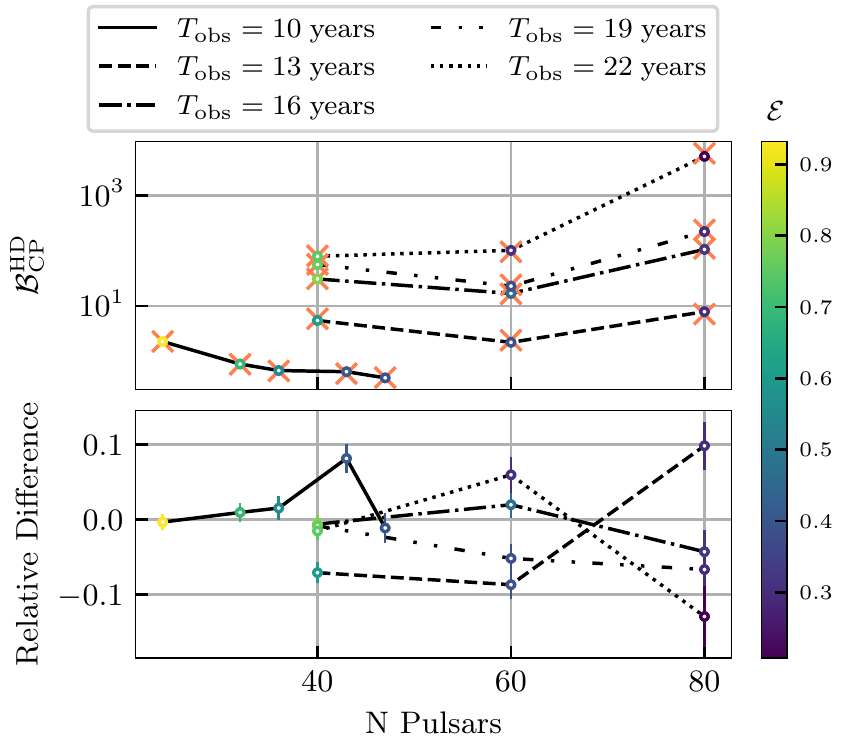}
    \caption{Top: $\BF$ vs number of pulsars for sets of fixed observation times between 10 and 22 years, $T_{\mathrm{obs}}$. Bayes factors recovered via reweighting, Eq.~\eqref{eq:BayesFactor}, are colored by their efficiency, $\mathcal{E}$, Eq.~\eqref{eq:Efficiency}. Bayes factors recovered via the hypermodel are plotted as coral $X$s. The error estimate of each point is described in the caption of Fig.~\ref{fig:error_BF}. 
    For a fixed number of pulsars, an increase in the observation time leads to a higher Bayes factor. In each case, as $N$  increases, the efficiency decreases. Additionally, as the observation time increases, efficiency tends to decrease, albeit less distinctly. 
    The relative difference between the direct sampling and reweighting Bayes factors remains quite small and is independent of both $T_\mathrm{obs}$ and $N$. The small errors and high efficiencies (each greater than $20\%$) imply that likelihood reweighting remains reliable when additional time and pulsars are added.}
    \label{fig:BayesFactor_vs_N_pulsar}
\end{figure}

In Fig.~\ref{fig:BayesFactor_vs_N_pulsar} we plot $\BF$ as a function of the number of pulsars, $N$, and for different observation durations, $T_\mathrm{obs}$. We find that the relative difference between the $\BF$ recovered by direct sampling and by reweighting remain within $10\%$ of each other, suggesting that the reweighting scheme remains valid for these extended data-sets. Moreover, we find that while the ratio between the HD and CP likelihood computation times is approximately constant across extended observation time, the ratio scales with the number of pulsars (ranging between 10 and 40). Thus as more pulsars are added to the dataset, reweighting becomes more important.

\section{Discussion and Conclusions}
\label{sec:discussion}

We have introduced a reweighting method to efficiently and reliably obtain GW posterior and marginal likelihood for a GWB model in PTA data analysis.
We first compute an inexpensive approximate posterior (CP) that omits pulsar-pulsar correlations, then reweight it to a full posterior (HD) that includes them.
We have validated this method by comparing reweighted posteriors and Bayes factors with distributions and factors obtained with direct sampling. Reweighting appears to be reliable and unbiased.
Even in cases with low reweighting efficiency (as defined by the reduction in the number of effective samples), the reweighted Bayes factor estimate remained robust up to $\BF > 10^6$, far larger than required for a confident GWB detection.

Even though our method requires evaluating the computationally expensive HD likelihood, it is still much more efficient than direct stochastic sampling. This is due to the additional evaluations required for the latter, which do not need to be repeated when reweighting.
Direct sampling results in very autocorrelated sample chains, which are thinned [by factors $N_{\mathrm{t}}\sim\mathcal{O}(10^3)$, on the order of the chain autocorrelation length] to obtain quasi-independent samples.
By contrast, reweighting is applied \emph{after} thinning, reducing the number of HD likelihood evaluations by $N_{\mathrm{t}}$.
In addition, the weights of Eq.\ \eqref{eq:likelihoodWeights} can be computed in parallel on multiple cores, allowing a further wall clock speedup (by the number $N_{\textrm{P}}$ of parallel processes).
Finally, if parallel tempering was used to sample the approximate model, only samples from the coldest chain should need be reweighted, decreasing the necessary number of computations by a factor of the number of chains $N_{\mathrm{c}}\sim \mathcal{O}(10)$. 

While the reweighting procedure is mathematically exact, the method is subject to sampling error; reweighted posteriors could have too few effective samples to accurately reflect the true distribution. Constructing generic diagnostic tools for such situations can be challenging, as the effective number of independent samples $\neff$ can vary between applications. In such cases, estimating the Bayes factor sampling error or inspecting posteriors visually can help identify undersampling.
If $\neff$ is low, a few strategies are available.
The simplest is to increase the number of samples for the approximate model.
A more sophisticated option involves the importance sampling of the approximate model by concentrating on the region of parameter space that the target seems to prefer.
In the most extreme case, so many approximate-model samples are needed that the method becomes less efficient than direct sampling. This happens when the efficiency drops to the ratio of likelihood computation times (e.g., to 1/25 for the NANOGrav 12.5yr dataset). If parallel tempering is used then ``hot'' chains, with a correspondingly broader posterior, could be used in situations where efficiency is low due to a lack of samples from the approximate distribution available to estimate tails in the target distribution.

The reweighting formalism is generic and can be applied to any pair of approximate and target distributions. For example, one could model a clock error by including a process with monopolar correlations in addition to the HD correlations. In this situation, extra parameters are added to the target model, which requires drawing samples from some proposal distribution for the new parameters (see~\cite{Romero-Shaw:2020aaj,Romero-Shaw:2020thy} for examples of reweighting between models with varying numbers of parameters). In practice, sampling error (efficiency) increases (decreases) if the approximate and target posteriors do not overlap, as quantified in Fig.~\ref{fig:KLDivergence} using the Kullback-Leibler divergence. 

Throughout this work we have presented examples that are based on the NANOGrav 12.5yr analysis. Although our simulations are consistent with the NANOGrav dataset and our understanding of the stochastic GWB, we have not simulated realistic radio frequency noise such as dispersion measure variations or solar wind fluctuations. More ``advanced'' noise modeling adds numerous extra parameters to each pulsar to measure chromatic effects~\cite{Tiburzi:2015kqa,Tiburzi:2019sal,Tiburzi:2020wuh,NANOGrav:2021yqt,Goncharov:2020krd} increasing the complexity of the analysis. Given that most of these additional parameters impact only individual pulsar measurements, a factorized-likelihood approach to estimate the CP model, followed by this reweighting scheme could significantly reduce the wall clock time of an analysis that uses more advanced noise models.

In the context of PTA searches for GWBs, the reweighting formalism introduced in this paper offers an accurate and computationally efficient shortcut to GW posteriors and HD vs. CP Bayes factors.
In this paper we tested the method on simulated datasets with increasing GWB amplitudes, which served a proxy for increased observing time and number of pulsars.
Our results suggest that reweighting remains robust for PTA datasets with Bayes factors of at least $10^6$, orders of magnitude larger than current results. Thus, our method can reliably characterize the GWB from PTA datasets for the foreseeable future and into the detection regime. 

\acknowledgements

We thank Ethan Payne for useful discussions about reweighting and Ken Olum, Steve Taylor, and Paul Baker for comments on the paper draft.
Numerical investigations were performed using services provided by the OSG Consortium~\cite{osg07, osg09}, which is supported by the National Science Foundation Grants No. 2030508 and No. 1836650. Additional computing resources were provided by Caltech's Theoretical AstroPhysics Including Relativity and Cosmology (TAPIR) group.
In addition to \texttt{enterprise}~\cite{enterprise, enterpriseExtensions}, our software stack included
\texttt{scipy}~\cite{2020SciPy-NMeth}, \texttt{matplotlib}~\cite{Hunter:2007}, \texttt{numpy}~\cite{harris2020array}, \texttt{pandas}~\cite{reback2020pandas}, and \texttt{corner}~\cite{corner}.
S.H. and P.M.M. acknowledge the VIPER PTA Summer School at Vanderbilt University, which was funded under NSF CAREER-2146016.
A.D.J. and K.C. acknowledge support from the Caltech and Jet Propulsion Laboratory President’s and Director’s Fund and the Sloan Foundation.
S.H. is supported by the National Science Foundation Graduate Research Fellowship under Grant No. DGE‐1745301.
P.M.M. and M.V. were supported by the NANOGrav Physics Frontiers Center, National Science Foundation (NSF), Grant No. 2020265.
Part of this research was carried out at the Jet Propulsion Laboratory, California Institute of Technology, under a contract with the National Aeronautics and Space Administration (80NM0018D0004).

\appendix

\bibliography{Refs.bib}

\end{document}